\documentstyle[preprint,aps,epsf,floats]{revtex}    
\def\ap#1#2#3   {{\em Ann. Phys. (NY)} {\bf#1} (#2) #3}
\def\apj#1#2#3  {{\em Astrophys. J.} {\bf#1} (#2) #3}
\def\apjl#1#2#3 {{\em Astrophys. J. Lett.} {\bf#1} (#2) #3}
\def\app#1#2#3  {{\em Acta. Phys. Pol.} {\bf#1} (#2) #3}
\def\ar#1#2#3   {{\em Ann. Rev. Nucl. Part. Sci.} {\bf#1} (#2) #3}
\def\cpc#1#2#3  {{\em Comp. Phys. Comm.} {\bf#1} (#2) #3}
\def\err#1#2#3  {{\it Erratum} {\bf#1} (#2) #3}
\def\ib#1#2#3   {{\it ibid.} {\bf#1} (#2) #3}
\def\jmp#1#2#3  {{\em J. Math. Phys.} {\bf#1} (#2) #3}
\def\ijmp#1#2#3 {{\em Int. J. Mod. Phys.} {\bf#1} (#2) #3}
\def\jetp#1#2#3 {{\em JETP Lett.} {\bf#1} (#2) #3}
\def\jpg#1#2#3  {{\em J. Phys. G.} {\bf#1} (#2) #3}
\def\mpl#1#2#3  {{\em Mod. Phys. Lett.} {\bf#1} (#2) #3}
\def\nat#1#2#3  {{\em Nature (London)} {\bf#1} (#2) #3}
\def\nc#1#2#3   {{\em Nuovo Cim.} {\bf#1} (#2) #3}
\def\nim#1#2#3  {{\em Nucl. Instr. Meth.} {\bf#1} (#2) #3}
\def\np#1#2#3   {{\em Nucl. Phys.} {\bf#1} (#2) #3}
\def\pcps#1#2#3 {{\em Proc. Cam. Phil. Soc.} {\bf#1} (#2) #3}
\def\pl#1#2#3   {{\em Phys. Lett.} {\bf#1} (#2) #3}
\def\prep#1#2#3 {{\em Phys. Rep.} {\bf#1} (#2) #3}
\def\prev#1#2#3 {{\em Phys. Rev.} {\bf#1} (#2) #3}
\def\prl#1#2#3  {{\em Phys. Rev. Lett.} {\bf#1} (#2) #3}
\def\prs#1#2#3  {{\em Proc. Roy. Soc.} {\bf#1} (#2) #3}
\def\ptp#1#2#3  {{\em Prog. Th. Phys.} {\bf#1} (#2) #3}
\def\ps#1#2#3   {{\em Physica Scripta} {\bf#1} (#2) #3}
\def\rmp#1#2#3  {{\em Rev. Mod. Phys.} {\bf#1} (#2) #3}
\def\rpp#1#2#3  {{\em Rep. Prog. Phys.} {\bf#1} (#2) #3}
\def\sjnp#1#2#3 {{\em Sov. J. Nucl. Phys.} {\bf#1} (#2) #3}
\def\spj#1#2#3  {{\em Sov. Phys. JEPT} {\bf#1} (#2) #3}
\def\spu#1#2#3  {{\em Sov. Phys.-Usp.} {\bf#1} (#2) #3}
\def\zp#1#2#3   {{\em Zeit. Phys.} {\bf#1} (#2) #3}
\def\trhozm{T_\rho^{(2 {m})}}
\def\trhodm{T_\rho^{(3 {m})}}
\def\trhovm{T_\rho^{(4 {m})}}

\begin{document}
\tighten
%
\preprint{
\font\fortssbx=cmssbx10 scaled \magstep2
\hbox to \hsize{
\hfill \vtop{   \hbox{\bf TTP96-48 }
                \hbox{\bf hep-ph/9702382\\}
                \hbox{February 1997}
                \hbox{}
                \hbox{              } }}
}
\title{Isospin Violation in $\tau\rightarrow 3\pi\nu_\tau$}
\author{E. Mirkes and R. Urech}
\address{Institut f\"ur Theoretische Teilchenphysik, 
         Universit\"at Karlsruhe,\\ D-76128 Karlsruhe, Germany\\[2mm]}
\maketitle
\thispagestyle{empty}
\begin{abstract}
Isospin violating signals in the $\tau^-\rightarrow(3\pi)^-\nu_\tau$ decay
mode are discussed. For the $\tau^-\rightarrow \pi^-\pi^-\pi^+\nu_\tau$
decay mode, isospin violation arises from the vector current contribution
in the $\tau^-\rightarrow \omega\pi^-\nu_\tau$ decay with the subsequent
isospin violating $\omega$ decay into $\pi^+\pi^-$. We demonstrate that
such effects may be observed in presently available data through the
measurement of the interference effects of these vector current
contributions with the dominating axial vector current, {\it i.e.}  through
a measurement of the structure functions $W_F,W_G,W_H$ and $W_I$. In the
case of the $\tau^- \to \pi^0 \pi^0 \pi^- \nu_\tau$ decay mode, a vector
current contribution is generated by $\eta \pi^0$ mixing in the decay chain
$\tau^- \to \eta \rho^- \nu_\tau \to \pi^0 \pi^0\pi^- \nu_\tau$.  We find
that this effect is rather small, the magnitude of the associated
interference terms being too low for present statistics.
\end{abstract}
%
\newpage
%

\section{Introduction}
Isospin rotations have been successfully used in $\tau$ decays into an even
number of final state pions to relate the vector current to the
corresponding cross sections measured in electron positron collisions
\cite{ks,eidelman}.  In the case of the two pion mode, the $\tau$ decay
rate has been measured with a relative error below one percent which is of
the size of possible isospin violating effects.  Isospin symmetry relations
are also very useful to relate various decay amplitudes in $3\pi\nu_\tau$,
$KK\pi\nu_\tau$ and $K\pi\pi\nu_\tau$ final states
\cite{pseudo2,rouge,colangelo}.

Isospin violation effects in the decay $\tau^-\to\omega\pi^-\nu_\tau$ have
been discussed in \cite{deckerm}.  Such signals could be revealed by an
analysis of the angular distribution in the $\omega\pi^-$ system.  Another
interesting isospin violating process is provided by the decay
$\tau^-\to\eta\pi^-\nu_\tau$. The different theoretical predictions for the
branching ratio \cite{pich87,neufeld95} are still one order of magnitude
smaller than the actual experimental upper limit \cite{cleo_limit96}.

In this article we will concentrate on possible isospin violating effects
in the $\tau\rightarrow 3 \pi\nu_\tau$ decay mode.  Although the
theoretical uncertainties in this decay mode are fairly large, observations
of small isospin violating effects (below 1\% to the rate) might be
possible with presently available statistics.  The sensitivity to such
small effects is provided by an analysis of angular distributions. The
relevant information is encoded in structure functions \cite{km1,km2} which
allow to reconstruct the form factors in the dominating axial current and
in the small isospin violating vector current contributions. In particular
the interference effects between the vector and the axial vector amplitudes,
given by the structure functions $W_F,W_G,W_H,W_I$ allow for such a
measurement.  Any nonvanishing contribution to these structure functions
would be a clear signal of isospin violation in the three pion decay mode
of the $\tau$.  After specifying the isospin violating vector form factor,
we will present numerical predictions for these structure functions
including the full dependence on the resonance structure.
We also analyze Dalitz distributions
for the purely axial vector structure functions.
Such distributions, in particular for the structure function $W_D$,
are fairly sensitive to the details of the $\rho$ sub-resonance 
implementation in the underlying models.

A branching fraction of $0.6\%$ in the $\tau\rightarrow 3\pi\nu_\tau$
mode due to isospin violation has been reported by the ARGUS collaboration
\cite{argus_michel}. Their analysis is based on a study of eight different
contributions to the amplitude. Unfortunately the relevant interference
terms with the isospin conserving part of the amplitude cannot be traced
out unambiguously from that work.

The paper is organized as follows: The general structure of the decay
amplitude and the structure function formalism in the three meson decay
mode is briefly summarized in Sec.~\ref{sec_amp} and a particular choice
for the form factors in the axial vector current, the K\"uhn--Santamaria
model \cite{ks}, is specified in Sec.~\ref{sec_ks}.  Isospin violating
contributions to an additional vector current form factor will be discussed
in Sec.~\ref{sec_iso1} (Sec.\ref{sec_iso2}) for the $\tau^-\rightarrow
\pi^-\pi^-\pi^+\nu_\tau$ ($\tau^-\rightarrow \pi^0\pi^0\pi^-\nu_\tau$)
decay mode.  The relevant hadronic matrix elements are determined in the
vector meson dominance model. We obtain as a by-product from the decay
$\tau^-\to\eta\pi^0\pi^-\nu_\tau$ a new parametrization for the transition
of the $\rho$ resonance into three pseudoscalar mesons which is also needed
as an input {\it e.g.} for the $\tau$ decay into $KK\pi\nu_\tau$ final
states. Finally, isospin violating signals induced by the vector current
form factor are discussed in Sec.~VI.

\newpage
\section{Three Meson Decay Modes:{\protect\\}
         Form Factors and Structure Functions}
\label{sec_amp}
The  matrix element ${\cal{M}}$  for the semi-leptonic $\tau$ decay 
into three mesons $h_1, h_2, h_3$
\begin{equation}
\tau(l,s)\rightarrow\nu_\tau(l^{\prime},s^{\prime})
+h_{1}(q_{1},m_{1})+h_{2}(q_{2},m_{2})+h_{3}(q_{3},m_{3}) 
\label{tau3h}
\end{equation}
can be expressed in the following form:
\begin{equation}
{\cal M}(\tau\rightarrow\nu_\tau\,\,h_1h_2h_3)=
\frac{G_F}{\sqrt{2}}\,\,\,
\bigl(^{\cos\theta_{c}}_{\sin\theta_{c}}\bigr)\,\,\,
\bar{u}(l^\prime,s^\prime)\gamma_{\mu}(1-\gamma_5)u(l,s)\,\,J^{\mu}.
\label{mdef}
\end{equation}
In Eq.~(\ref{mdef})
$G_F$ denotes the Fermi-coupling constant and   $\theta_c$ is the 
Cabibbo angle. 
The hadronic current 
%
%
\begin{equation}
J^{\mu}(q_1,q_2,q_3)=\langle h_{1}(q_{1})h_2(q_2)h_3(q_3))
|V^{\mu}(0)-A^{\mu}(0)|0\rangle 
\end{equation}
is  characterized by four independent form factors $F_1,F_2,F_3,F_4$
\cite{km1}.
These form factors are in general functions of 
$s_1=(q_2+q_3)^2, s_2=(q_1+q_3)^2$, $s_3=(q_1+q_2)^2$
and $Q^2$, which is conveniently chosen as an additional variable.
\begin{eqnarray}
J^{\mu}(q_{1},q_{2},q_{3})
&=&     V_{1}^{\mu}\,F_{1}
    + V_{2}^{\mu}\,F_{2}
    +\,i\, V_{3}^{\mu}\,F_{3}   
    + V_{4}^{\mu}\,F_{4} 
    \label{f1234}
\end{eqnarray}
with
\begin{equation}
\begin{array}{ll}
V_{1}^{\mu}&= (q_{1}-q_{3})_{\nu}\,T^{\mu\nu}  \\
V_{2}^{\mu}&= (q_{2}-q_{3})_{\nu}\,T^{\mu\nu}  \\
V_{3}^{\mu}&= \epsilon^{\mu\alpha\beta\gamma}q_{1\,\alpha}q_{2\,\beta}
                                             q_{3\,\gamma} 
\\
V_{4}^{\mu}&=q_{1}^{\mu}+q_{2}^{\mu}+q_{3}^{\mu}\,=Q^{\mu} .
    \end{array}
\label{videf}
\end{equation}
$T^{\mu\nu}$ denotes the transverse projector
\begin{equation}
T_{\mu\nu}=  g_{\mu \nu} - \frac{Q_\mu Q_\nu}{Q^2}  \>.
\label{trans}
\end{equation}
$F_1$ and $F_2$ determine the spin one component of the axial vector
current induced amplitude, $F_4$ the spin zero part which is given by the
matrix element of the divergence of the axial vector current. The vector
current induced amplitude is responsible for the form factor $F_3$.  All
form factors may contribute in the general three meson case
\cite{kaons,pseudo1,pseudo2}.  $G$-parity conservation and PCAC in the
three pion decay mode implies $F_3=F_4=0$. However, isospin violation is
expected to give a nonvanishing contribution to $F_3$.
Such contributions will be studied 
in the last three sections of this paper.

The three meson decay in Eq.~(\ref{tau3h}) is  most easily analyzed in the
hadronic rest frame $\vec{q_1}+\vec{q_2}+\vec{q_3}=0$.
The orientation of the hadronic
system is in general  characterized by 
three Euler angles ($\alpha,\beta$ and $\gamma$) as introduced in
\cite{km1,km2}.
Of particular interest are the distributions of the normal
to the Dalitz plane and the distributions around this normal.
Performing the analysis in the hadronic rest frame has the advantage that
the product of the hadronic tensor
($H^{\mu\nu}=J^{\mu}(J^{\nu})^{\dagger}$) and the leptonic tensor reduces
to a sum $ L^{\mu\nu}H_{\mu\nu}=\sum_{X} \bar{L}_XW_X$.  The leptonic
factors $\bar{L}_X$ factorize the dependence on the Euler angles.  For the
definition of these angles and the explicit dependence of the coefficients
$\bar{L}_X$ on $\alpha,\beta$ and $\gamma$ see ref.~\cite{km1}.  The (in
general 16) hadronic structure functions $W_{X}$ correspond to 16 density
matrix elements for a hadronic system in a spin one and spin zero state
(nine of them originate from a pure spin one state and the remaining
originate from a pure spin zero state or from interference terms between
spin zero and spin one).  These structure functions contain the dynamics of
the three meson decay and depend only on the form factors $F_i$ and on the
hadronic invariants $Q^2$ and the Dalitz plot variables $s_{i}$.  The
scalar contribution is expected to be small \cite{scalar} for all three
meson final states and will be neglected in the subsequent discussion of
this paper\footnote{Using an ansatz for a scalar contribution in the
$3\pi\nu_\tau$ decay mode as specified in \cite{km1}, U. M\"uller
constrained such a contribution in the branching ratio to be less than 0.84
\% by analyzing the spin-zero-spin-one structure functions with 1994 OPAL
data \cite{ute_beta}. Note that a possible scalar contribution would not
contribute to the vector-axial vector interference structure functions in
Eq.~(\ref{wi}) which are important for an observation of isospin violating
effects.}.  Instead of the 16 real structure functions which characterize
the general hadronic tensor $H^{\mu\nu}$ one thus deals only with nine
functions $W_X$.  These nine structure functions can be divided in four
functions which arise only from the axial vector current ($W_{A,C,D,E}$),
one from the vector current ($W_B$) and the remaining four from the
interference of the axial vector and vector current ($W_{F,G,H,I}$).  The
latter will be of particular importance in the subsequent discussion.

The dependence of the structure functions on the form factors $F_i$
reads \cite{km1}:\\
{\underline{Axial vector structure functions:}}
\begin{eqnarray} 
W_{A}  &=&   \hspace{3mm}(x_{1}^{2}+x_{3}^{2})\,|F_{1}|^{2}
           +(x_{2}^{2}+x_{3}^{2})\,|F_{2}|^{2}  +2(x_{1}x_{2}-x_{3}^{2})\,
           \mbox{Re}\left(F_{1}F^{\ast}_{2}\right)
                                   \nonumber \\
W_{C}  &=&  \hspace{3mm} (x_{1}^{2}-x_{3}^{2})\,|F_{1}|^{2}
           +(x_{2}^{2}-x_{3}^{2})\,|F_{2}|^{2}    +2(x_{1}x_{2}+x_{3}^{2})\,
           \mbox{Re}\left(F_{1}F^{\ast}_{2}\right)
                                \label{walldef}   \\
W_{D}  &=&  \hspace{3mm}2\left[ x_{1}x_{3}\,|F_{1}|^{2}
           -x_{2}x_{3}\,|F_{2}|^{2}\right.      \left.  +x_{3}(x_{2}-x_{1})\,
           \mbox{Re}\left(F_{1}F^{\ast}_{2}\right)\right]
                                   \nonumber \\
W_{E}  &=& -2x_{3}(x_{1}+x_{2})\,\mbox{Im}\left(F_{1}
                    F^{\ast}_{2} \right) \nonumber
\end{eqnarray}
{\underline{Vector structure function:}}
\begin{eqnarray}
W_{B}  &=& \hspace{3mm} x_{4}^{2}|F_{3}|^{2}
\label{wb}
\end{eqnarray}
{\underline{Axial vector--vector interference structure functions:}}
\begin{eqnarray}  \hspace{3mm}
W_{F}  &=&  \hspace{3mm}
          2x_{4}\left[x_{1}\,\mbox{Im}\left(F_{1}F^{\ast}_{3}\right)
                     + x_{2}\,\mbox{Im}\left(F_{2}F^{\ast}_{3}\right)\right]
                                   \nonumber \\[1mm]
W_{G}  &=&- 2x_{4}\left[x_{1}\,\mbox{Re}\left(F_{1}F^{\ast}_{3}\right)
                     + x_{2}\,\mbox{Re}\left(F_{2}F^{\ast}_{3}\right)\right]]
                                   \nonumber \\[1mm]
W_{H}  &=& \hspace{3mm}
      2x_{3}x_{4}\left[\,\mbox{Im}\left(F_{1}F^{\ast}_{3}\right)
                     -\,\mbox{Im}\left(F_{2}F^{\ast}_{3}\right)\right]
                                   \label{wi} \\[1mm]
W_{I}  &=&- 2x_{3}x_{4}\left[\,\mbox{Re}\left(F_{1}F^{\ast}_{3}\right)
                     -\,\mbox{Re}\left(F_{2}F^{\ast}_{3}\right)\right]
                                   \nonumber 
\end{eqnarray}
The variables $x_i$ are defined by
   \begin{eqnarray}
x_{1}&=& V_{1}^{x}=q_{1}^{x}-q_{3}^{x}\nonumber\\[1mm]
x_{2}&=& V_{2}^{x}=q_{2}^{x}-q_{3}^{x}\nonumber\\[1mm]
x_{3}&=& V_{1}^{y}=q_{1}^{y}=-q_{2}^{y}\\[1mm] 
x_{4}&=& V_{3}^{z}=\sqrt{Q^{2}}x_{3}q_{3}^{x}\nonumber
   \end{eqnarray}
where $q_i^{x}$ ($q_i^{y}$) denotes
the $x$ ($y$) component of the momentum of
meson $i$ in the hadronic rest frame.
They can easily be expressed in terms of $s_1$, $s_2$ and $s_3$
\cite{km1}.
$W_A(Q^2,s_i)$ and $W_B(Q^2,s_i)$
govern the rate and the distributions in the Dalitz plot through
\begin{equation}
\Gamma(\tau\rightarrow 3h\nu_\tau)=
           \frac{G^{2}}{12m_\tau} 
\bigl(^{\cos\theta_{c}}_{\sin\theta_{c}}\bigr)^2
\frac{1}{(4\pi)^5}
\int \frac{dQ^{2}}{Q^4} ds_1 ds_2 \,(m_\tau^{2}-Q^{2})^{2}\,
          \left( 1+\frac{2Q^2}{m_\tau^2}\right)\,
           \,\,\left(W_{A}+W_{B}\right)
\label{rate}
\end{equation}
The remaining structure functions determine the angular distribution.
All of them can be determined by a measurement of the $\beta$ and $\gamma$
dependence even without reconstructing the $\tau$ rest frame.

\section{Axial Vector Current Contribution to 
         $\tau^-\rightarrow (3\pi)^-\nu_\tau$}
\label{sec_ks}
$\tau$ decays into three pions are dominated by the axial vector current
which allows for significant simplifications: $G$-parity implies $F_3=0$,
Bose symmetry relates $F_1$ and $F_2$ through
$F_2(s_1,s_2,Q^2)=F_1(s_2,s_1,Q^2)$ and PCAC leads to $F_4=0$. Note that
the structure functions $W_{B,F,G,H,I}$ in Eqs.~(\ref{wb},\ref{wi}) vanish
for $F_3=0$.

The two like-sign pions in
$\tau^-\rightarrow \pi^-\pi^-\pi^+\nu_\tau$ and
$\tau^-\rightarrow \pi^0\pi^0\pi^-\nu_\tau$ 
are labeled such that
$|\vec{p}_2| > |\vec{p}_1|$ and $p_3$ refers to the unlike-sign pion.
The normalization of the form factors $F_1$ and $F_2$ for the three pion
decay mode is determined in the chiral limit\footnote{We 
use the Condon-Shortley phase conventions.} 
\cite{fischer80},
\begin{equation}
F_1=F_2=i \frac{2\sqrt{2}}{3f_{\pi}},\hspace{2cm}
f_\pi = 93\mbox{ MeV}
\end{equation}
\begin{figure}[t]            
\begin{picture}(0,0)(0,0)
\includegraphics{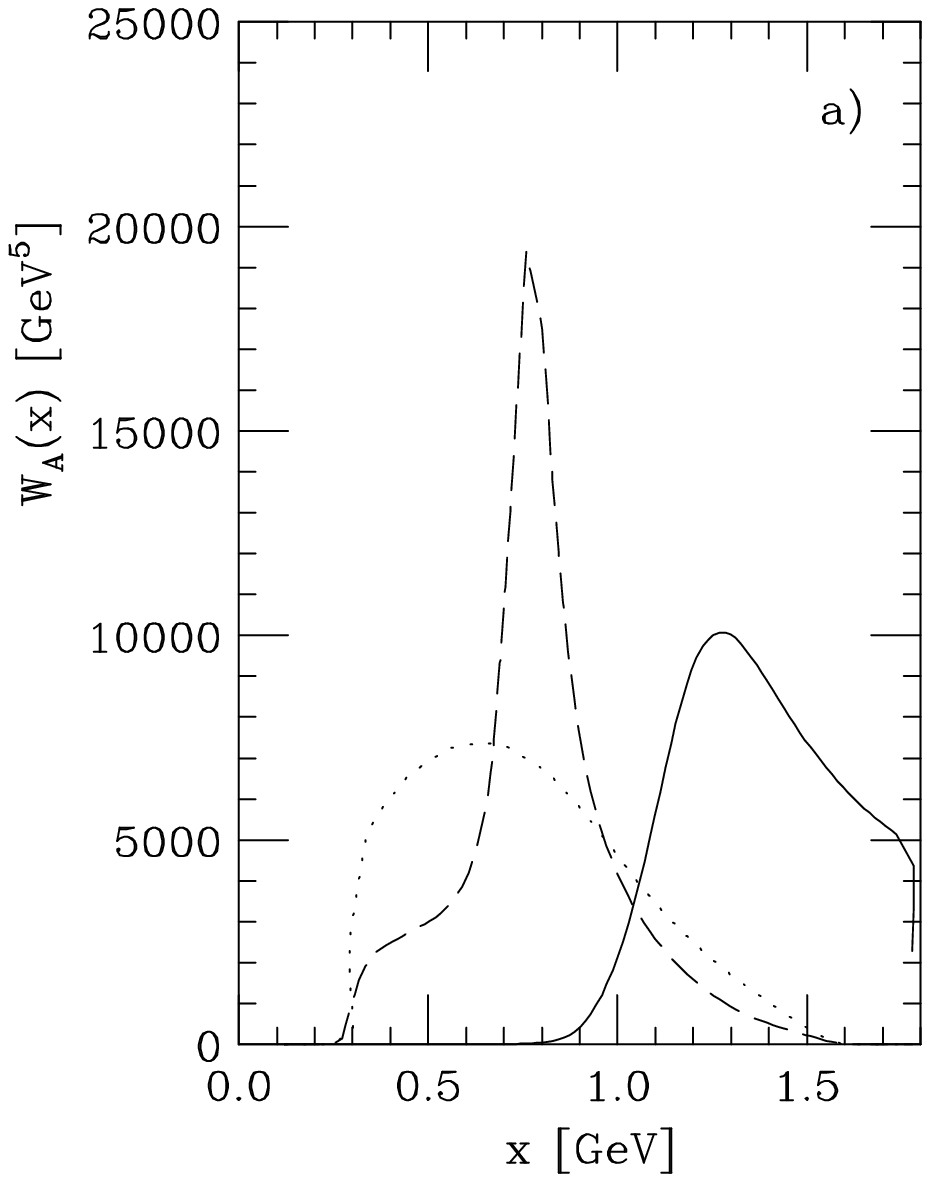}
\includegraphics{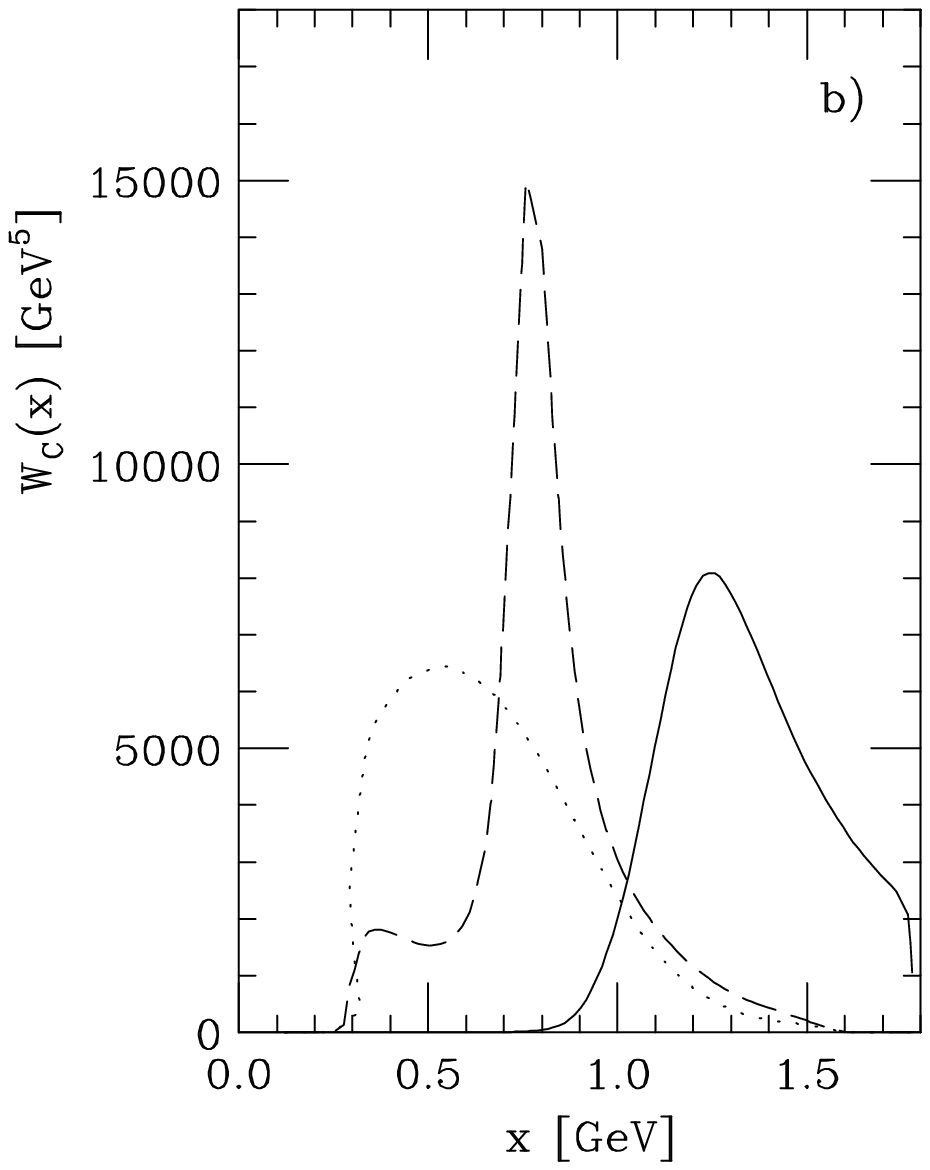}
\includegraphics{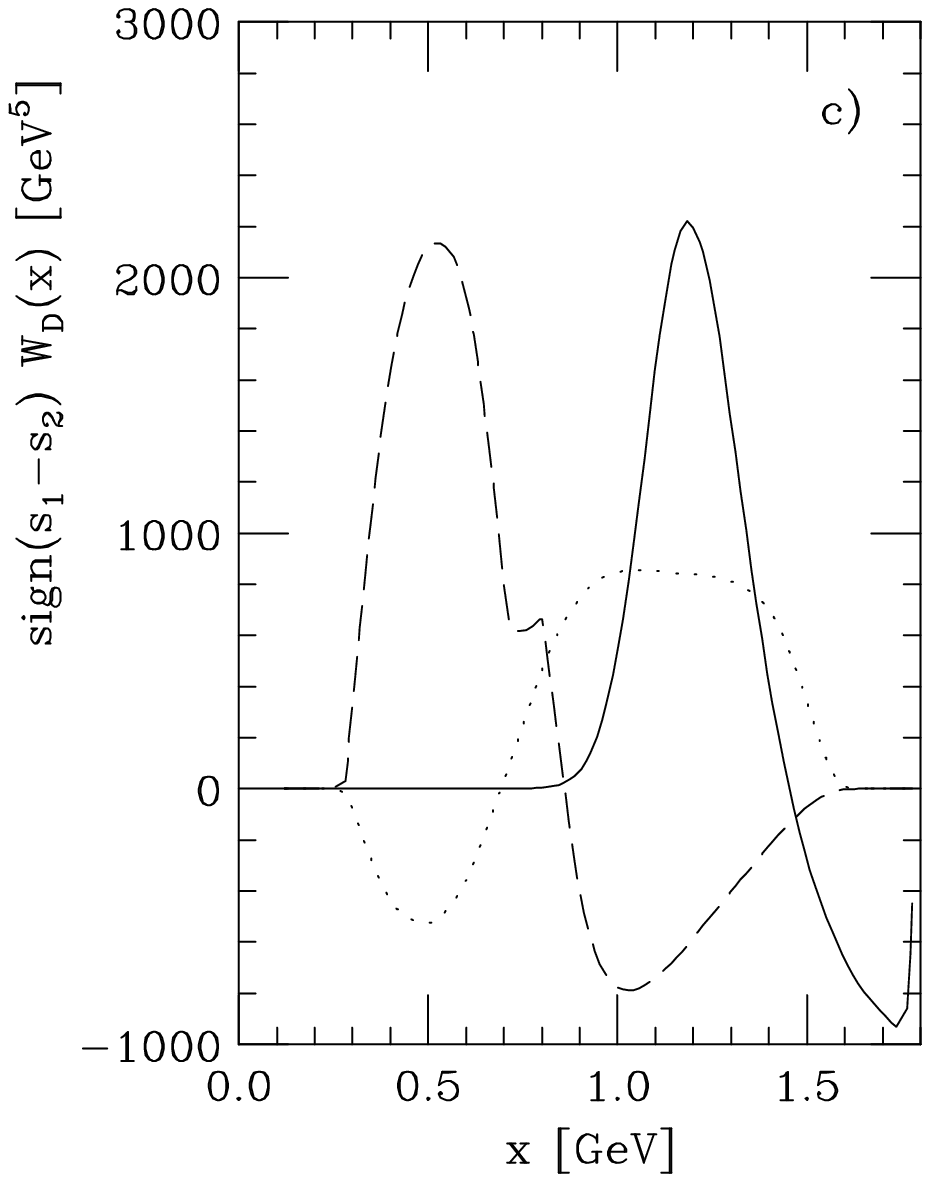}
\includegraphics{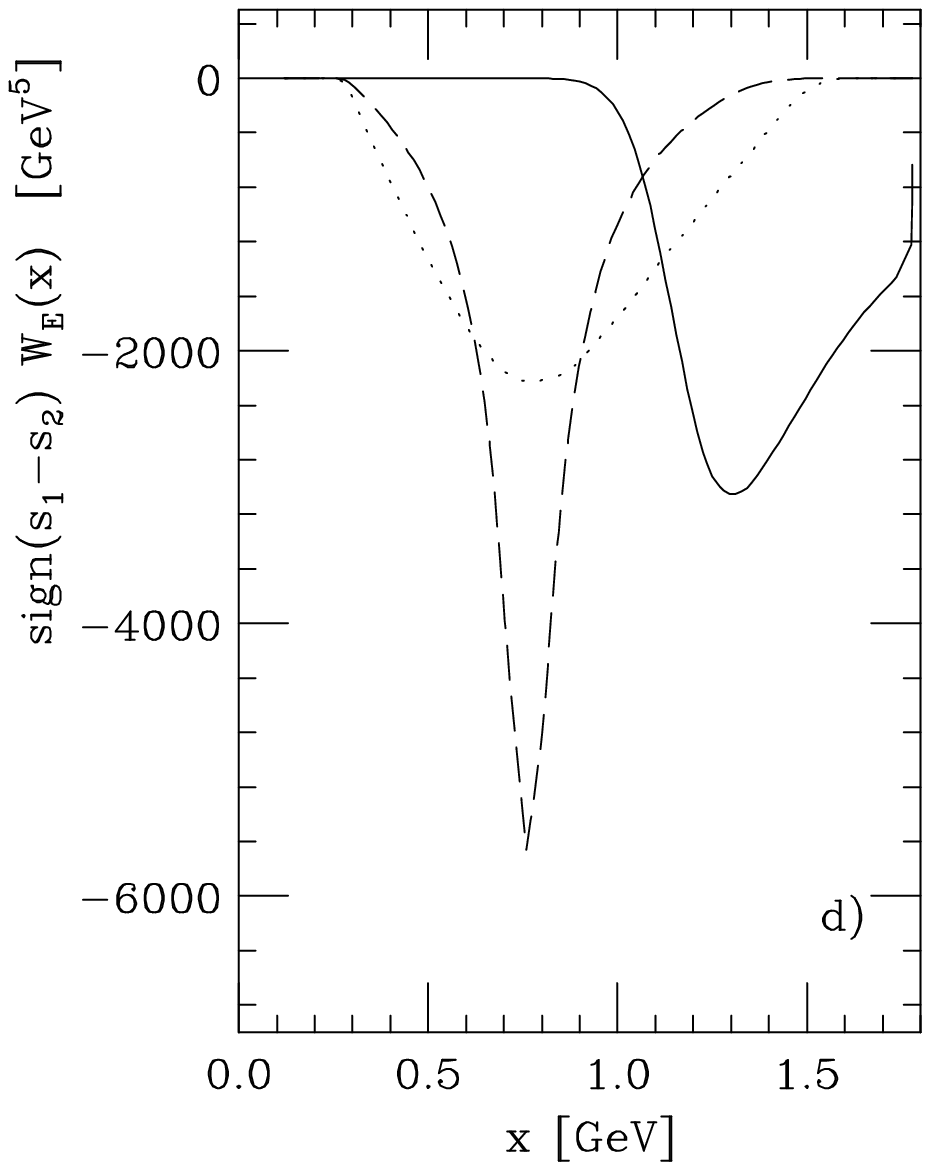}
\end{picture}
\vspace{14.5cm}
\caption{Invariant mass distributions
$x=\protect\sqrt{Q^2}=m(\pi^-\pi^-\pi^+)$   (solid),
$x=\protect\sqrt{s_{1,2}}=m(\pi^+\pi^-)$ (dashed) and
$x=\protect\sqrt{s_3}=m(\pi^-\pi^-)$ (dotted) of the structure functions
$W_A$ (a), $W_C$ (b), $sign(s_1-s_2)W_D$ (c), $sign(s_1-s_2)W_E$ (d) in the
$\tau^-\rightarrow(3\pi)^-\nu_\tau$ decay mode.  }
\label{fig_wacde}
\end{figure}
For large $Q^2,s_1$ and $s_2$ these form factors are modulated by
resonances in the $3\pi$ and $2\pi$ channel. Following the ansatz of K\"uhn
and Santamaria \cite{ks}, one has
\begin{eqnarray}
F_1(Q^2,s_2)&=& i \frac{2\sqrt{2}}{3f_{\pi}}\,\,\,
BW_{a_1}(Q^2)\,\trhozm(s_2) \\
F_2(Q^2,s_1)&=& i \frac{2\sqrt{2}}{3f_{\pi}}\,\,\,
BW_{a_1}(Q^2)\,\trhozm(s_1)
\end{eqnarray}
The Breit--Wigner functions $\mbox{BW}_X(s)$ are parametrized including
energy dependent widths,
\begin{equation}
  \mbox{BW}_{X}(s) = 
  \frac{m_{X}^2} {m_{X}^2 - s - i \sqrt{s} \Gamma_{X}(s)}\quad,
  \hspace{2cm}\mbox{BW}_X(0)=1
\end{equation}
For the $a_1$ we have in particular
\begin{equation}
\Gamma_{a_1}(s)=\frac{m_{a_1}}{\sqrt{s}}\,\Gamma_{a_1}\,
\frac{g(s)}{g(m_{a_1}^2)}
\hspace{1cm}
m_{a_1} = 1.251 \,\, \mbox{GeV}\>,  \hspace{1cm}
\Gamma_{a_1 } = \, 0.475\,\, \mbox{GeV}
\end{equation}
%
%
%
%
where the function $g(s)$ has been calculated in \cite{ks} and is derived
from the observation, that the axial vector resonance $a_1$ decays
predominately into three pions.

The superscript $(2m)$ in the $\rho$ form factor $\trhozm(s)$ denotes the
subsequent decay into two pions. In the parametrization of $\trhozm(s)$ one
allows for a contribution of the first excitation $\rho'$,
\begin{eqnarray} 
   \trhozm(s) & = & \frac{1}{1 + \beta_\rho}
  \Big[ \mbox{BW}_\rho(s) + \beta_\rho\, \mbox{BW}_{\rho'}(s)
  \Big] \>,
   \label{beta}
\end{eqnarray}
with the energy dependent width 
\begin{equation}
\Gamma_\rho(s)=\Gamma_\rho\,\frac{m_\rho^2}{s}
\left(\frac{s-4m_\pi^2}{m_\rho^2-4m_\pi^2}\right)^{3/2}
\end{equation}
and similarly for the $\rho^\prime$. The parameters are given by 
\begin{eqnarray}
\beta_\rho = -0.145\>, & & \nonumber \\
m_\rho = 0.773 \, \mbox{GeV}\>, & & \Gamma_\rho = 0.145 \, \mbox{GeV}\>,
\nonumber \\
m_{\rho'} = 1.370 \, \mbox{GeV}\>, & & \Gamma_{\rho'} = 0.510 \, \mbox{GeV}\>.
 \label{eqnrho}
\end{eqnarray}
which have been determined from $e^+ e^- \to \pi^+ \pi^-$ in
\cite{ks}. 
Predictions for the $(s_1-s_2)$-integrated structure functions 
$w_{X}(Q^2)=\int d s_1 d s_2 W_X(Q^2,s_1,s_2)$ for
$X=A,C,D,E$ based on this model 
are in good agreement with data \cite{exp3pi}.
The invariant $3\pi$ and $2\pi$ mass distributions for the four integrated
nonvanishing structure functions $W_{A}, W_C, sign(s_1-s_2) W_D,
sign(s_1-s_2)W_E$ in Fig.~\ref{fig_wacde} reveal the importance of the
$a_1$ (solid) and $\rho$ (dashed) resonances.  The $\sqrt{s}_{3}$
distribution (dotted line) is then fixed by phase space restrictions and
the $\sqrt{Q^2}$ and ${\sqrt{s}_{1,2}}$ distributions through $s_3 = Q^2 -
s_1 -s_2 +3 m_\pi^2$.  The structure functions $W_D$ and $W_E$ are combined
with an energy ordering $sign(s_1-s_2)$ to account for Bose symmetry.  The
$\sqrt{s}_{1,2}$ distributions of $W_A, W_C$ and $sign(s_1-s_2)W_F$ have a
clear peak around the $\rho$ resonance, whereas
$sign(s_1-s_2)W_D(\sqrt{s}_{1,2})$ has a surprisingly different behaviour
in the K\"uhn-Santamaria model. The distribution shows a relatively wide
peak around $\sqrt{s}_{1,2}=0.5$ GeV and only a much smaller additional
peak around the $\rho$ mass.  In contrast, the $\sqrt{s}_{1,2}$
distribution for $sign(s_1-s_2)W_D$ based on the model in \cite{isgur} has
its maximum around the $\rho$ mass and only a small additional peak around
$\sqrt{s}_{1,2}=0.5$.  An experimental confirmation of the predictions for
the $\sqrt{s}_{1,2}$ distributions in the axial vector structure functions
shown in Fig.~\ref{fig_wacde} and in particular in $sign(s_1-s_2)W_D$ would
be a good test of the details in the $\rho$ resonance structure in the
K\"uhn Santamaria model which we use for the two axial vector form factors
$F_1$ and $F_2$.

\section{Vector Current Contribution  to $\tau^-\to \pi^-\pi^-\pi^+\nu_\tau$}
\label{sec_iso1}
More detailed studies, such as testing the magnitude of amplitudes induced
by $F_3$ through isospin violation, are possible and will be discussed in
the following. Since they affect the angular distributions through
interference terms between the (small) contribution from $F_3$ with the large
contributions from $F_1$ and $F_2$ (see Eq.~(\ref{wi})), 
they should be accessible in measurements of the
structure functions $W_{F,G,H,I}$, already with the statistics of ongoing
experiments.

A small vector current contribution ($\sim F_3$ in Eq.~(\ref{f1234})) to
the $\tau^- \to \pi^-\pi^-\pi^+\nu_\tau$ mode is expected to arise from the
$\tau^-\to\omega\pi^-\nu_\tau$ decay with the subsequent isospin violating
$\omega$ decay into $\pi^+\pi^-$.  $G$-parity requires that the
$\omega\pi^-$ system is in a $1^{-}$ state and hence the
$\tau^-\to\omega\pi^-\nu_\tau$ decay can only proceed via a vector current.
The hadronic matrix element is determined through \cite{pseudo2,omegapi1}
(for another approach see \cite{omegapi2})
\begin{eqnarray}
\langle \omega(\tilde{q}_{1},\lambda)\pi^-(\tilde{q}_{2})| 
V^{\mu}(0)|0\rangle &=& i\,\epsilon^{\mu\alpha\beta\gamma}\, 
                        \varepsilon^{\ast}_\alpha (\tilde{q}_{1},\lambda)\,
                        \tilde{q}_{1\,\beta}\tilde{q}_{2\,\gamma}\,
                        F_V^{(\omega\pi)}(Q^2) 
                        \nonumber \\
F_V^{(\omega\pi)}(Q^2) &=& i\,\frac{f_{\rho^-}g_{\rho\omega\pi}}
                           {m^2_{\rho}}\, 
\label{omegapi}                           \trhovm(Q^2)
\\
Q^2 &=& (\tilde{q}_1 + \tilde{q}_2)^2 \nonumber
\end{eqnarray}
where $V^\mu$ is the vector part of the weak current. Note that we have
fixed the sign of $F_V^{(\omega\pi)}(0)$ from $\pi^0 \to \gamma
\gamma$. $f_{\rho^-}$ is the coupling of the charged $\rho^\pm$ to the
gauge boson $W^\pm$ and is related to the
$\rho^0\gamma$ coupling $f_{\rho}$, $g_{\rho\omega\pi}$ is measured in the
decays $\omega \to \pi^0 \gamma$ and $\omega \to \pi^+ \pi^-
\pi^0$\cite{pdg96}, respectively,
\begin{eqnarray}
f_{\rho^-} &=& \sqrt{2}\,f_{\rho}
           \;\simeq\; 0.17 \,\mbox{ GeV}^2 \nonumber \\
g_{\rho\omega\pi} &=& \Bigg\{  
\begin{array}{ll}
11.7 \pm 0.4 \,\mbox{ GeV}^{-1}\hspace{1cm} \mbox{from}
&  \Gamma \left(\omega \to \pi^0 \gamma \right)\\
15.0 \pm 0.1 \,\mbox{ GeV}^{-1} 
\hspace{1cm}\mbox{from}&  
\Gamma \left(\omega \to \pi^+ \pi^- \pi^0 \right) 
\end{array}
\end{eqnarray}
\setlength{\unitlength}{0.7mm}
\begin{figure}[t]             
\begin{picture}(150,165)(-70,-85)
\mbox{\epsfxsize10.0cm\epsffile[78 222 480 650]{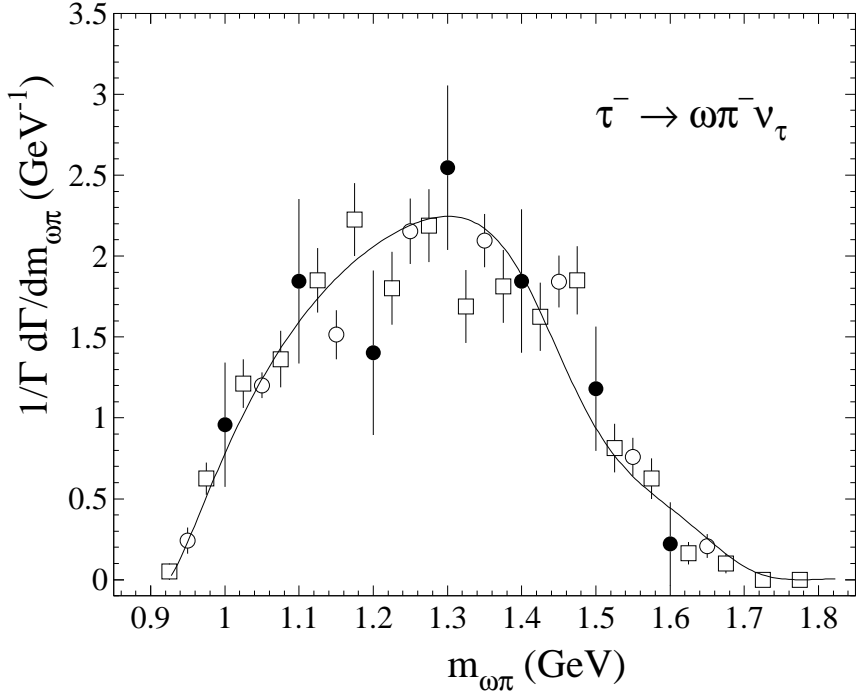}}
\end{picture}
\caption{The invariant $\omega\pi$ mass spectrum from
$\tau^-\to\omega\pi^-\nu_\tau$ measured by ARGUS
\protect\cite{argus_omegapi} (filled circles), by CLEO
\protect\cite{cleo_omegapi} (opened circles) and by ALEPH
\protect\cite{aleph_omegapi} (opened squares). The solid line shows the fit
result to Eqs.~(\protect\ref{omegapi},\protect\ref{bwomegapi}).}
\label{opi}
\end{figure}
The $\rho$-meson and its radial excitations are possible resonance
candidates for the vector form factor $\trhovm(s)$, where the superscript
$(4m)$ refers to the (anomalous) VMD decay chain $\rho \to \omega \pi \to 4
\pi$. The admixture of the radial excitations in $\trhovm(s)$ is expected
to differ from the corresponding $\rho$ form factor $\trhozm(s)$ with a
dominant two pion decay in Eq.~(\ref{beta}). Here we allow for an admixture
of the $\rho^\prime$ and the $\rho^{\prime\prime}$ via
\begin{eqnarray}
   \trhovm (s) & = & \frac{1}{1 + \lambda + \kappa}
  \Big[ \mbox{BW}_\rho(s) + \lambda \mbox{BW}_{\rho'}(s)
  + \kappa \mbox{BW}_{\rho''}(s)\Big] 
\label{bwomegapi}
\end{eqnarray}
where we fix the parameters to the PDG \cite{pdg96} values, which yields
\begin{eqnarray}
\begin{array}{ll}
m_\rho = 0.773 \mbox{ GeV}\>,    &\Gamma_\rho = 0.145 \mbox{ GeV}\> \\ 
m_{\rho'} = 1.465 \mbox{ GeV}\>, &\Gamma_{\rho'} = 0.310 \mbox{ GeV}\>\\ 
m_{\rho''} = 1.70 \mbox{ GeV}\>, &\Gamma_{\rho''} = 0.235 \mbox{ GeV}\>.
\end{array}
\label{masses}
\end{eqnarray}
The parameters $\lambda$ and $\kappa$ are obtained from a fit to the
normalized invariant mass spectrum of $\tau^-\to\omega\pi^-\nu_\tau$ data
\cite{argus_omegapi,cleo_omegapi,aleph_omegapi}, see Fig.~\ref{opi},
\begin{eqnarray}
\lambda &=& - 0.054 \pm 0.012 \nonumber \\
\kappa  &=& - 0.036 \pm 0.004  \label{fitopi}\\
\chi^2/\rm{d.o.f.} &=& 53.3/31 \quad \nonumber.
\end{eqnarray}
Note that the errors should be taken as an educated guess only, since we
fit to the published data with the correlation matrices to be diagonal, and
the mass and width parameters in Eq.~(\ref{masses}) are considered 
exact values. The values in Eq.~(\ref{fitopi}) lead to the following
branching ratios, depending strongly on the $\omega$ decay channel from
which one extracts $g_{\rho\omega\pi}$,
\begin{eqnarray}
\cal B \left(\tau^- \to \omega\pi^-\nu_\tau \right) &=& \Bigg\{  
\begin{array}{ll}
\left( 0.98 \pm 0.21 \right) \,\% \hspace{1cm} 
& g_{\rho\omega\pi}=11.7 \pm 0.4 \,\mbox{ GeV}^{-1} \\
\left( 1.61 \pm 0.23 \right) \,\%              
&   g_{\rho\omega\pi}=15.0 \pm 0.1 \,\mbox{ GeV}^{-1} 
\end{array} 
\end{eqnarray}
The errors in the branching ratios are dominated by the errors
in $\lambda$ and $\kappa$ in Eq.~(\ref{fitopi}).
A comparison to the measured experimental branching ratios shows that small
values for $g_{\rho\omega\pi}$ are excluded,
\begin{equation}  
{\cal B}_{exp.} \left(\tau^- \to \omega\pi^-\nu_\tau \right) = 
\left (1.92 \pm 0.08 \right)\,\%
\end{equation}
where we combined the measured branching fractions from CLEO
\cite{cleo_omegapi} and ALEPH \cite{aleph_omegapi} . Thus we will put
$g_{\rho\omega\pi} = 15.0 \,\mbox{ GeV}^{-1}$ in the following, keeping in
mind that the measured $\tau$ decay rate would even require a higher
value of $g_{\rho\omega\pi}$.

The transition $\omega\to \pi^+ \pi^-$ is assumed to proceed through
$\rho^0 \omega$ mixing and is written in the form
\begin{equation}\label{omegapipi}
\langle \pi^+(k_1) \pi^-(k_2) | {\cal T} | \omega(k,\lambda) \rangle =
 \frac{\theta_{\rho\omega} g_{\rho\pi\pi}}{m^2_\rho}\mbox{BW}_\rho(k^2)
(k_1 - k_2)_\mu \,\varepsilon^\mu(k,\lambda) 
\end{equation}
where $g_{\rho\pi\pi}$ is related to the decay $\rho^0 \to \pi^+ \pi^-$,
and the $\rho^0\omega$ mixing parameter $\theta_{\rho\omega}$ is measured
in $e^+ e^- \to \pi^+ \pi^-$ experiments \cite{thomas95},
\begin{equation}
g_{\rho\pi\pi} = 6.08\quad,\hspace{2cm}
\theta_{\rho\omega} = ( - 3.97 \pm 0.20 )\times 10^{-3} \mbox{ GeV}^2 
\quad .
\end{equation} 
Combining the amplitudes in Eqs. (\ref{omegapi},\ref{omegapipi}) one
obtains for the three pion decay mode after summation over the polarization
$\lambda$ of the intermediate $\omega$ state,
\begin{eqnarray}
\langle \pi^- \pi^- \pi^+|V^\mu|0 \rangle = \sum_\lambda \,\langle \pi^+
\pi^- | {\cal T} | \omega(p,\lambda) \rangle \langle \omega(p,\lambda)\,
\pi^- |V^{\mu}|0\rangle \;\frac{-\,1}{m_\omega^2} \mbox{BW}_\omega(s)
\end{eqnarray}
where $s=p^2$ is the momentum transfer and the width in
$\mbox{BW}_\omega(s)$ is chosen to be energy independent due to its
smallness,
\begin{equation}
BW_{\omega}(s)= 
\frac{m^2_{\omega}}{m^2_{\omega} - s - i m_{\omega}
\Gamma_{\omega}},
\hspace{1cm}
M_{\omega}= 0.782 \mbox{ GeV},
\hspace{1cm}
\Gamma_{\omega}= 8.4\mbox{ MeV}
\end{equation}
With the identity $\sum_\lambda \varepsilon_\mu (p,\lambda)
\varepsilon^\ast_\nu(p,\lambda) = - g^{\mu\nu} + p^\mu p^\nu/m^2$ we find
the following parametrization of the form factor $F_3$,
\begin{eqnarray}
\langle \pi^-(q_1) \pi^-(q_2) \pi^+(q_3)|V^\mu(0)|0 \rangle   
&=& i\,\epsilon^{\mu\alpha\beta\gamma}\,q_{1\,\alpha}q_{2\,\beta}
    q_{3\,\gamma}\,F_3(s_1,s_2,Q^2) \nonumber \\
F_3(s_1,s_2,Q^2) 
&=& - 2 \frac{\theta_{\rho\omega} g_{\rho\pi\pi}}{m^2_\rho m^2_\omega}
F_V^{(\omega\pi)}(Q^2) \nonumber \\
&& \times \left[ \mbox{BW}_\omega(s_1)\mbox{BW}_\rho(s_1) 
   - \mbox{BW}_\omega(s_2)\mbox{BW}_\rho(s_2)  \right]\quad .
\label{f3omegapi}
\end{eqnarray}
%
\section{Vector Current Contribution  to $\tau^-\to \pi^0\pi^0\pi^-\nu_\tau$}
\label{sec_iso2}
In the case of the $\tau^- \to \pi^0 \pi^0 \pi^- \nu_\tau$ decay mode we
assume that the vector current contribution is generated by $\eta \pi^0$
mixing in the decay chain $\tau^- \to \eta \rho^- \nu_\tau \to \pi^0
\pi^0\pi^- \nu_\tau$.

The $\tau^- \to \eta \pi^0 \pi^-\nu_\tau$ decay is allowed to proceed in
the Standard Model via a vector current induced by the Wess-Zumino anomaly
part in the Lagrangian \cite{wzw}. A normalization of the form factor
$F_3^{(\eta\pi\pi)}$ is fixed in the chiral limit and a parametrization of
$F_3^{(\eta\pi\pi)}$ reads \cite{pich87,pseudo1,gomez90,braaten87}
\begin{eqnarray}
\langle \eta (q_1) \pi^0 (q_2) \pi^-(q_3)|V^\mu(0)|0 \rangle 
&=& i\,\epsilon^{\mu\alpha\beta\gamma}\,q_{1\,\alpha}q_{2\,\beta}
    q_{3\,\gamma}\,F_3^{(\eta\pi\pi)}(s_1,Q^2) \nonumber \\
F_3^{(\eta\pi\pi)}(s_1,Q^2) 
&=& i\,\frac{\sqrt{6}}{12 \pi^2 f_\pi^3}\,
    \trhodm(Q^2)\, \trhozm(s_1) \quad .
\end{eqnarray}
\setlength{\unitlength}{0.7mm}
\begin{figure}[t]             
\begin{picture}(150,165)(-70,-85)
\mbox{\epsfxsize10.0cm\epsffile[78 222 480 650]{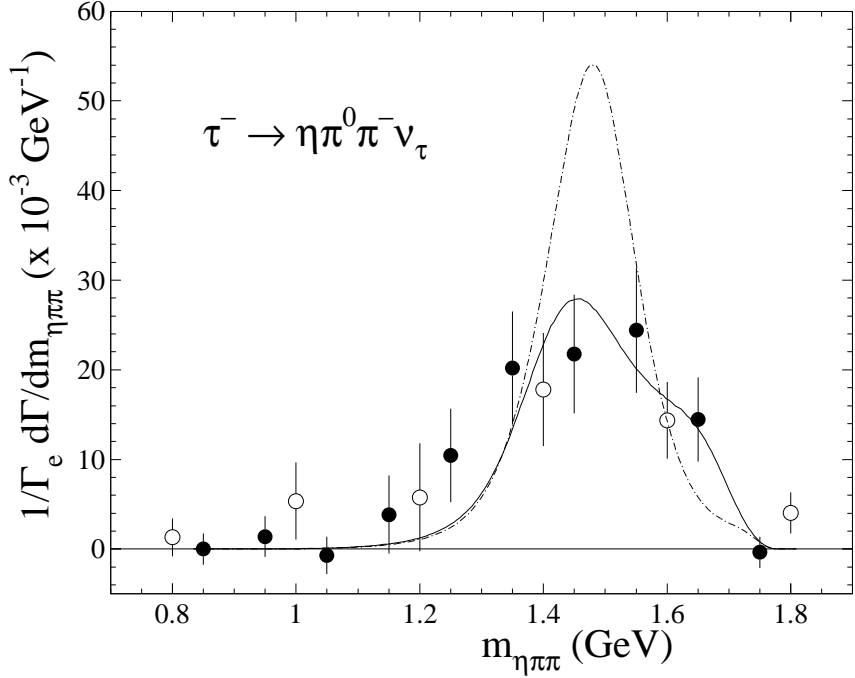}}
\end{picture}
\caption{The $\eta\pi\pi$ mass spectrum from $\tau^- \to \eta \pi^0\pi^-
\nu_\tau$ measured by CLEO \protect\cite{cleo_etapipi} (filled circles) and
by ALEPH \protect\cite{aleph_omegapi} (open circles) normalized to
$\Gamma_e = \Gamma(\tau^- \to e^- \overline{\nu}_e \nu_\tau)$. The solid
line shows the fit result to Eq.~(\protect\ref{bwetapipi}), the dashed line
represents the $\eta\pi\pi$ mass spectrum obtained from
$e^+e^-\rightarrow\eta\pi\pi$ data \protect\cite{gomez90,DM2}.}
\label{etapipi}
\end{figure}
The form and parameters of $\trhozm(s)$ are given in
Eqs.~(\ref{beta},\ref{eqnrho}). For the form factor $\trhodm(s)$ the
superscript $(3m)$ implies the anomalous transition $\rho \to \eta \rho \to
\eta \pi \pi$. In \cite{pseudo2,pseudo1,gomez90}, a form for $\trhodm(s)$
including $\rho,\rho^\prime$ and $\rho^{\prime\prime}$ was used, which has
been obtained from a fit to (fairly poor) $e^+e^-\rightarrow\eta\pi\pi$
data \cite{gomez90,DM2}.  However, new measurements for $\tau^- \to \eta
\pi^0\pi^-\nu_\tau$ have become available allowing now for a direct
determination of $\trhodm(s)$ in $\tau$ decays
\cite{aleph_omegapi,cleo_etapipi}. A direct fit to the differential decay
rate for $\tau^-\rightarrow\eta\pi^0\pi^-\nu_\tau$ normalized to $\Gamma_e
= \Gamma(\tau^- \to e^- \overline{\nu}_e \nu_\tau)$ as shown in
Fig.~\ref{etapipi} (solid line) yields for the coefficients $\xi$ and
$\sigma$:
\begin{eqnarray}
   \trhodm(s) &=& \frac{1}{1 + \xi + \sigma}
  \Big[ \mbox{BW}_\rho(s) + \xi \mbox{BW}_{\rho'}(s)
  + \sigma \mbox{BW}_{\rho''}(s)\Big] \nonumber \\
\xi &=& - 0.22 \pm 0.03 \nonumber \\
\sigma &=& - 0.10 \pm 0.01 \nonumber \\
\chi^2/\rm{d.o.f.} &=& 11.0/14 \quad .
\label{bwetapipi}
\end{eqnarray}
where the masses and widths of the resonances are given in
Eq.~(\ref{masses}). Again the errors have to be considered educated
ones, see the remark in Sec.~III. The branching fraction that we obtain is
compatible with the measured decay rate,
\begin{eqnarray}
{\cal B} \left(\tau^- \to \eta \pi^0 \pi^- \nu_\tau \right) 
&=& \left( 0.14 \pm 0.05 \right)\% \nonumber \\
{\cal B}_{exp.} \left(\tau^- \to \eta \pi^0 \pi^- \nu_\tau \right) 
&=& \left( 0.17 \pm 0.03 \right)\% \quad .
\end{eqnarray}
where we give the weighted average of the experimental branching fractions
from CLEO \cite{cleo_etapipi} and ALEPH \cite{aleph_omegapi}. 
Thus the invariant mass distribution and the decay rate for the
$\tau^-\rightarrow \eta\pi^0\pi^-\nu_\tau$ decay mode are well
described by these parameters. On the other hand we found that the
$\eta\pi^0\pi^-$ invariant mass spectrum in Fig.~\ref{etapipi}
is only poorly described by the $\trhodm(s)$ parametrization
based on the $e^+e^-\rightarrow\eta\pi\pi$ data (dashed line
in Fig.~\ref{etapipi}).

For the isospin violating form factor in the three pion decay we deduce the
form
\begin{eqnarray}
\langle \pi^0(q_1) \pi^0(q_2) \pi^-(q_3)|V^\mu(0)|0 \rangle 
&=& i\,\epsilon^{\mu\alpha\beta\gamma}\,q_{1\,\alpha}q_{2\,\beta}
    q_{3\,\gamma}\,F_3(s_1,s_2,Q^2) \nonumber \\
F_3(s_1,s_2,Q^2) 
&=&  \varepsilon \, \left[ F_3^{(\eta\pi\pi)}(s_1,Q^2)
                     - F_3^{(\eta\pi\pi)}(s_2,Q^2) \right]\quad ,
\label{f3eta}
\end{eqnarray}
where an estimate of the $\eta \pi^0$ mixing parameter $\varepsilon$ is
given by \cite{gasser82}
\begin{equation}
\varepsilon = (1.05 \pm 0.07) \times 10^{-2} \quad . 
\end{equation}

An additional decay channel would be induced by $\eta\eta'$ mixing with a
subsequent $\eta' \pi^0$ transition. Experimentally, the decay $\tau^- \to
\eta' \pi^0 \pi^- \nu_\tau$ has not been observed \cite{shelkov} and thus a
reliable parametrization of the associated form factor
$F_3^{(\eta'\pi\pi)}$ is missing.
We therefore neglect possible contributions
from the $\eta'$ as an intermediate state.

\section{Numerical Results}
After having fixed our model for the isospin violating
vector current contributions to the three pion decay mode,
we next discuss numerical effects of this contribution
to the  decay widths and
in particular to the 
structure functions $W_B,sign(s_1-s_2)W_F,sign(s_1-s_2)W_G,W_H$ and $W_I$.
The structure functions $W_F$ and $W_G$ are again combined
with an energy ordering $sign(s_1-s_2)$ to account for Bose symmetry.

\begin{figure}[bt] 
\vspace*{2cm}           
\begin{picture}(0,0)(0,0)
\includegraphics{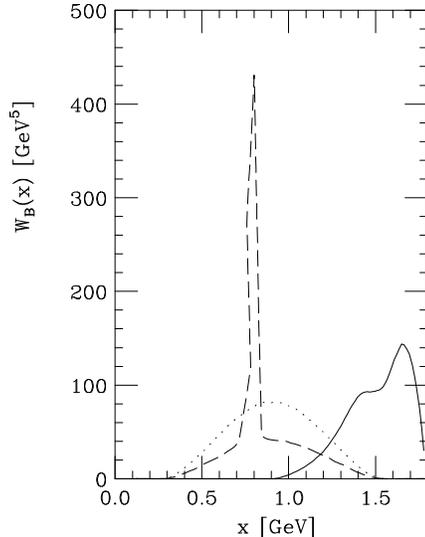}
\end{picture}
\vspace{7cm}
\caption{ Invariant mass distributions
$x=\protect\sqrt{Q^2}=m(\pi^-\pi^-\pi^+)$ (solid),
$x=\protect\sqrt{s_1}=\protect\sqrt{s_2}=m(\pi^+\pi^-)$ (dashed) and
$x=\protect\sqrt{s_3}=m(\pi^-\pi^-)$ (dotted) of the structure function
$W_B$ in the $\tau^-\rightarrow \pi^- \pi^- \pi^+\nu_\tau$ decay mode.  }
\label{fig_wb}
\end{figure}
\begin{figure}[t!]            
\vspace*{2cm}
\begin{picture}(0,0)(0,0)
\includegraphics{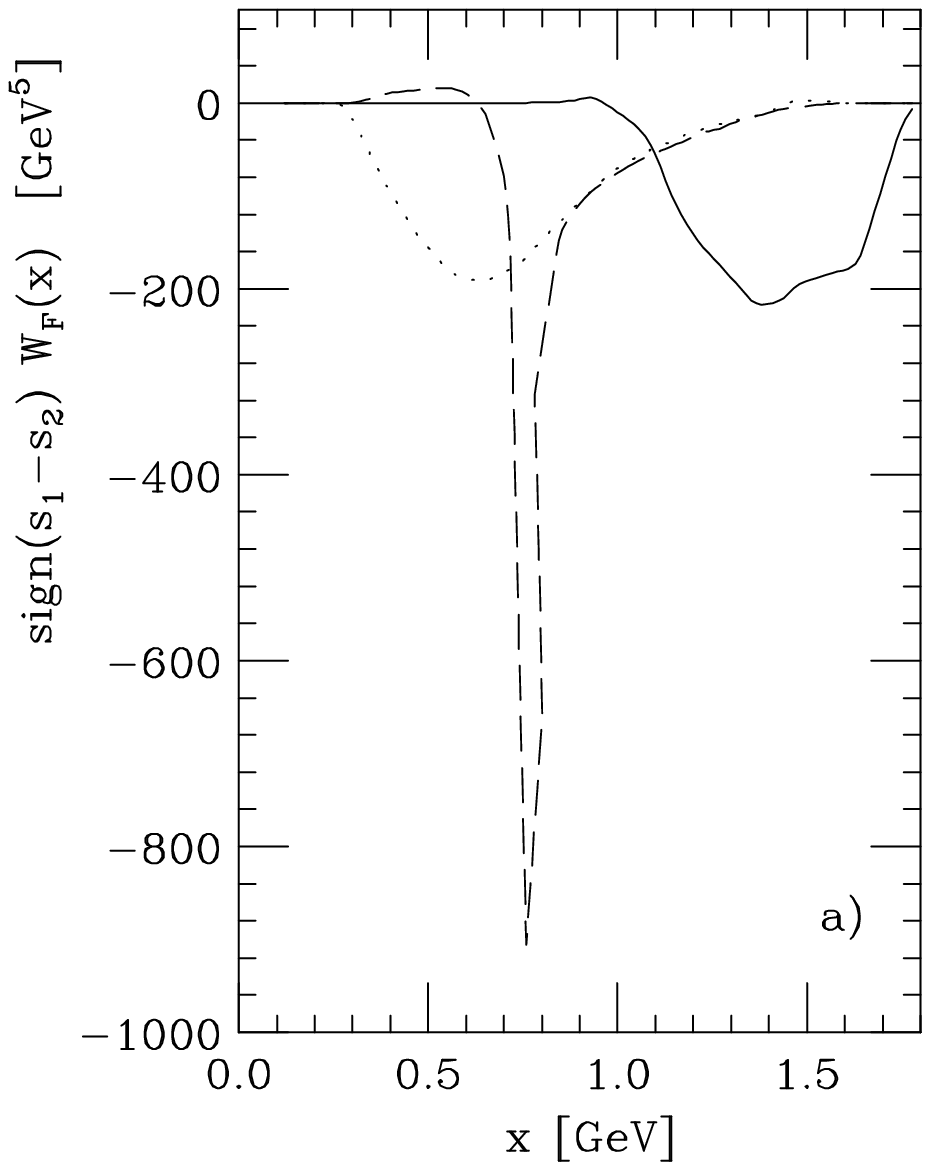}
\includegraphics{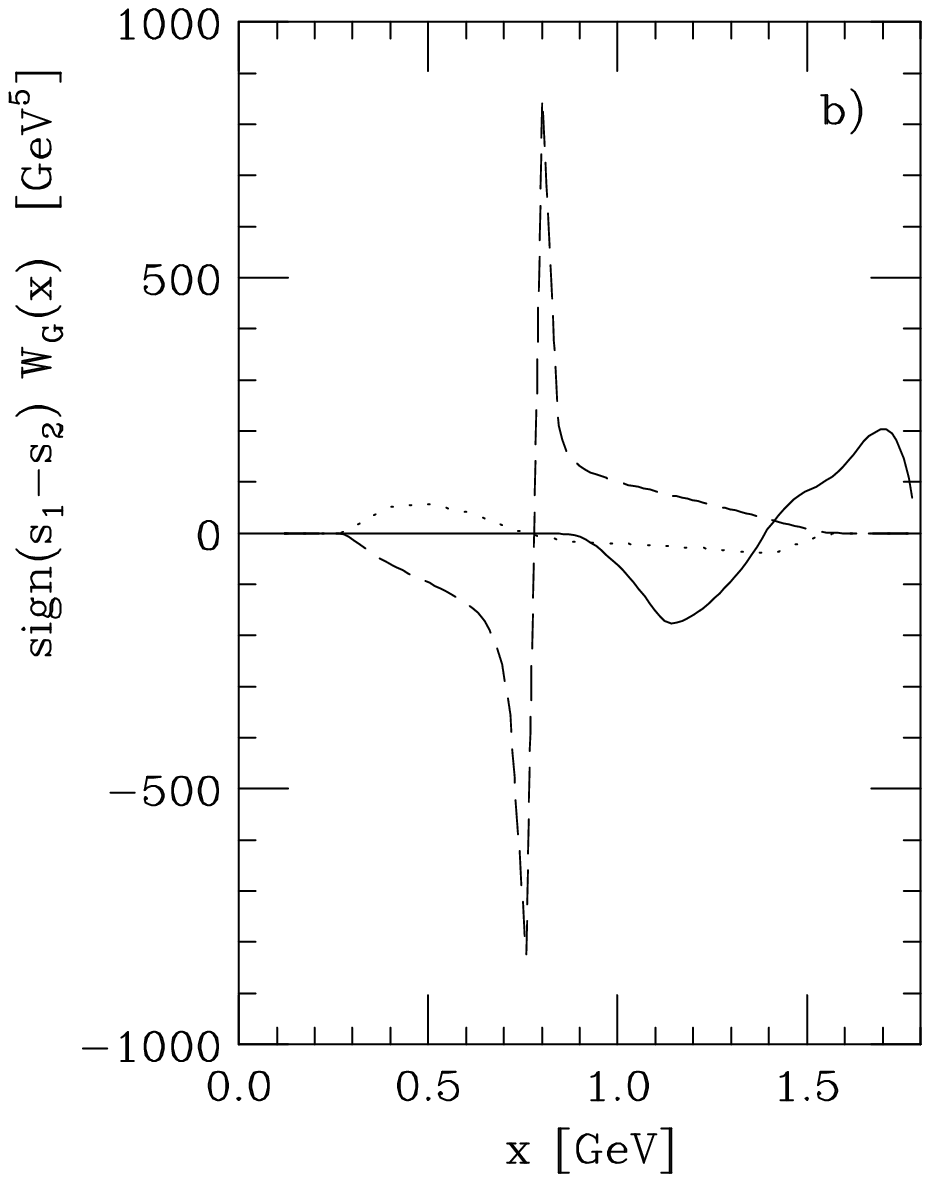}
\includegraphics{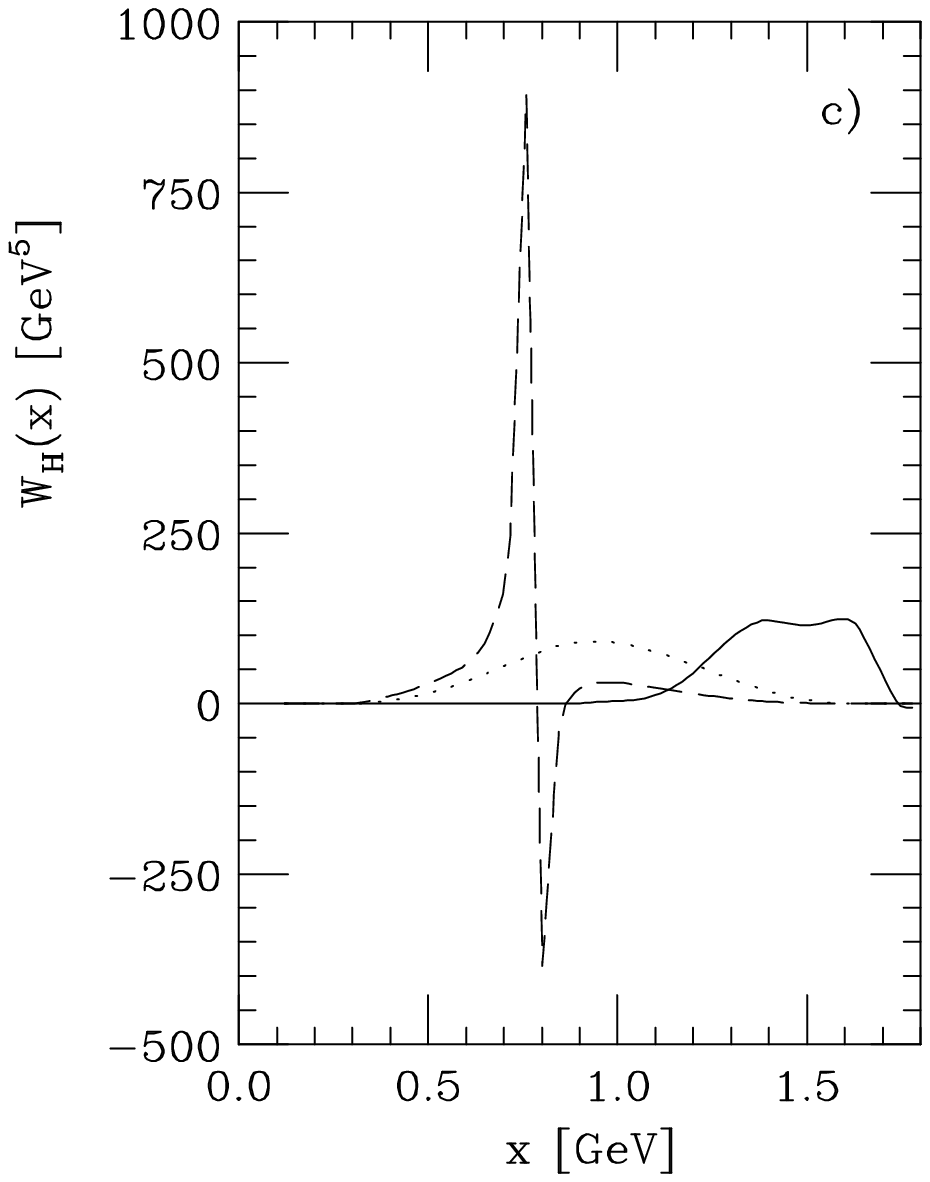}
\includegraphics{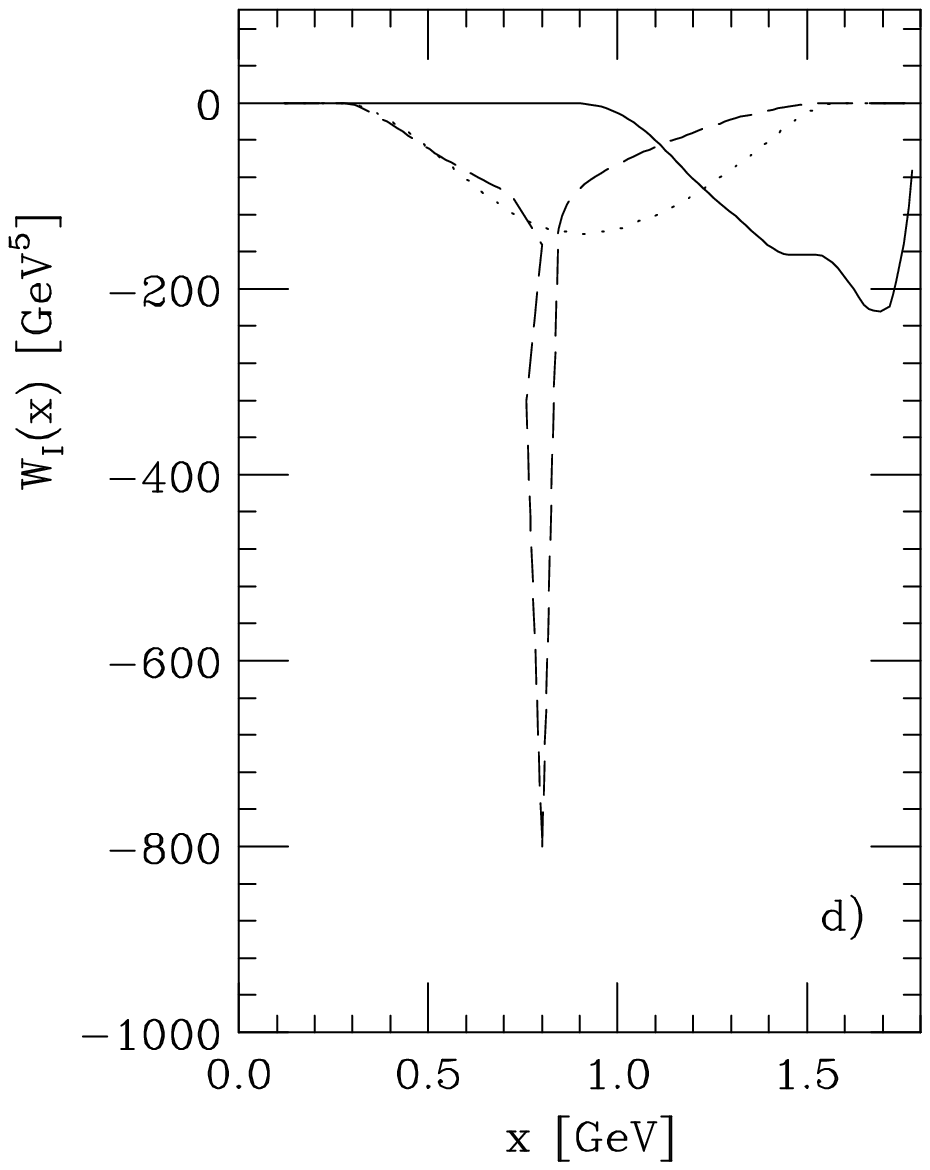}
\end{picture}
\vspace{14.5cm}
\caption{ Invariant mass distributions
$x=\protect\sqrt{Q^2}=m(\pi^-\pi^-\pi^+)$ (solid),
$x=\protect\sqrt{s_{1,2}}=m(\pi^+\pi^-)$ (dashed) and
$x=\protect\sqrt{s_3}=m(\pi^-\pi^-)$ (dotted) of the structure functions
$sign(s_1-s_2)W_F$ (a), $sign(s_1-s_2)W_G$ (b), $W_H$ (c), $W_I$ (d) in the
$\tau^-\rightarrow \pi^- \pi^- \pi^+\nu_\tau$ decay mode.  }
\label{fig_wfghi}
\end{figure}
Let us start with the $\tau^-\rightarrow\pi^-\pi^-\pi^+\nu_\tau$ decay
mode.  Fig.~\ref{fig_wb} shows the resonance structure of the pure vector
structure function $W_B$.  The $(\pi^-\pi^-\pi^+)$ mass distribution (solid
line) is dominated by the two higher radial excitations $\rho^\prime(1465)$
and $\rho^\prime(1700)$ of the $\rho$ resonance in Eq.~(\ref{omegapi}). The
narrow peak in the $\sqrt{s_1}=\sqrt{s_2}=m(\pi^+\pi^-)$ distribution
(dashed line) shows the dominance of the $\omega$ sub-resonance in the
vector current.  The shape of the $\sqrt{s_3}$ distribution (dotted line)
is fixed by phase space restrictions and the $\sqrt{Q^2}$ and 
${\sqrt{s_1}}, {\sqrt{s_2}}$
distributions through $s_3 = Q^2 - s_1 -s_2 +3 m_\pi^2$.
A comparison of $W_B$ in Fig.~\ref{fig_wb} with $W_A$ in Fig.~\ref{fig_wacde}
shows that the contribution to the decay rate from
$W_B$ is small compared to the axial vector structure function
$W_A$.
In fact, using Eq.~(\ref{rate}) we find that $W_B$ 
contributes  numerically $0.4\%$ to the decay rate, which is
slightly below the branching fraction that has been reported by ARGUS
\cite{argus_michel}, namely $0.6\%$. 
Due to the  large uncertainties in the axial vector part,
in particular in the $a_1$ width, isospin violating
effects cannot be seen by a rate measurement.
One could try to disentangle the structure functions
$W_A$ and $W_B$ by analyzing the difference in the
$\cos\beta$ distribution (see \cite{km1,km2}) ($\beta$ denotes
the angle between  the normal of the three pion plane and
the direction of the laboratory in the hadronic rest frame).
However, the sensitivity to the difference in the $\beta$ distribution
for these two structure functions is fairly small \cite{ute_beta}
and such an analysis is probably not possible with the current
statistics.

Much more promising is an analysis of the vector current contribution
through the measurement of the interference effects between the vector
current contribution with the dominating axial vector current contribution,
{\it i.e.}  through a measurement of the structure functions
$sign(s_1-s_2)W_F,sign(s_1-s_2)W_G,W_H$ and $W_I$.  The three meson and two
meson invariant mass distributions for these structure functions are shown
in Fig.~\ref{fig_wfghi}.  The shape of the three pion invariant mass
distributions (solid lines) is determined by the interference of the $a_1$
resonance in the form factors $F_1$ and $F_2$ and the $\trhovm$ resonance
in Eqs.~(\ref{omegapi},\ref{bwomegapi}).  The $\rho^\prime$ and
$\rho^{\prime\prime}$ peaks are visible in all four structure functions.
Similarly, the narrow peaks around 800 MeV in the $m(\pi^+\pi^-)$ invariant
mass distribution is a consequence of the interfering $\rho$ resonance in
$F_1$ and $F_2$ with the product of
$\mbox{BW}_{\omega}(s_{1,2})\mbox{BW}_\rho(s_{1,2})$ in $F_3$ as described
in Eq.(\ref{f3omegapi}).  The structure functions $sign(s_1-s_2)W_F$ and
$W_H$ are the most promising candidates to extract a vector current
contribution in an unambiguously way. An additional scalar contribution of
the same size as the vector current contribution discussed before would
contribute to two additional structure functions whose angular coefficients
are similar to those of $W_G$ and $W_I$ \cite{km1}, and thus a separation
of the vector current and such scalar effects might be very difficult with
presently available statistics in the data (see also \cite{ute_beta}). On
the other hand, the angular distributions which determine the structure
functions $sign(s_1-s_2)W_F$ and $W_H$ differ considerably from those
originating from possible spin-zero-spin-one interference effects. Any
nonvanishing contribution to these structure functions would therefore be a
clear signal of isospin violation.

In the decay $\tau^-\rightarrow\pi^0\pi^0\pi^-\nu_\tau$ we find the effects
of isospin violation to be negligibly small. Indeed, the contribution to
the decay rate from $W_B$ is of the the order $10^{-3}\%$. Those the
amplitudes in the invariant mass distributions of $W_B$ are very small when
compared to the corresponding invariant mass spectra in the decay
$\tau^-\rightarrow\pi^-\pi^-\pi^+\nu_\tau$. Even the distributions in the
interference terms do not have significant amplitudes. We therefore
conclude that isospin violation in the decay
$\tau^-\rightarrow\pi^0\pi^0\pi^-\nu_\tau$ can hardly be measured in
presently available data.\\[1cm]
To summarize: 
An isospin violating vector form factor is expected to give a
contribution of about 0.4\% to the decay rate in the
$\tau^-\rightarrow\pi^-\pi^-\pi^+\nu_\tau$ decay mode.
Sizable interference effects of this vector form factor with the
dominating axial vector form factors are discussed in detail.
These effects could be observed with presently
available statistics without reconstructing the $\tau$
rest frame. Any nonvanishing contribution to the corresponding
structure functions would be a clear signal of isospin violation. 
The corresponding signals in the
$\tau^-\rightarrow\pi^0\pi^0\pi^-\nu_\tau$ decay
are found to be considerably smaller.

\acknowledgements\noindent
We thank J.H. K\"uhn for helpful discussions.
%
%

\end{document}